\documentclass{aastex}          
\usepackage{spr-astr-addons}    
\usepackage{url}

\RequirePackage{color}

\usepackage{graphicx}
\usepackage[colorlinks=true,citecolor=blue,linkcolor=blue]{hyperref}

\begin{document}

\title{The Chinese space millimeter-wavelength VLBI array - a step toward imaging the most compact astronomical objects}
\shorttitle{Chinese space mm-wavelength VLBI array}
\shortauthor{Hong et al.}

\author{Xiaoyu~Hong\altaffilmark{1}, Zhiqiang~Shen\altaffilmark{1}, Tao~An\altaffilmark{1}, Qinghui~Liu\altaffilmark{1}, and the Chinese Space VLBI Array team} \affil{Shanghai Astronomical Observatory, Chinese Academy of Sciences, 80 Nandan Road, 200030 Shanghai, P.R. China, Email: xhong@shao.ac.cn, zshen@shao.ac.cn, antao@shao.ac.cn, liuqh@shao.ac.cn }

\begin{abstract}
The Shanghai Astronomical Observatory (SHAO) of the Chinese Academy of Sciences (CAS) is studying a space VLBI (Very Long Baseline Interferometer) program. The ultimate objective of the program is to image the immediate vicinity of the supermassive black holes (SMBHs) in the hearts of galaxies with a space-based VLBI array working at sub-millimeter wavelengths and to gain ultrahigh angular resolution. To achieve this ambitious goal, the mission plan is divided into three stages. The first phase of the program is called Space Millimeter-wavelength VLBI Array (SMVA) consisting of two satellites, each carrying a 10-m diameter radio telescope into elliptical orbits with an apogee height of 60000 km and a perigee height of 1200 km. The VLBI telescopes in space will work at three frequency bands, 43, 22 and 8 GHz. The 43- and 22-GHz bands will be equipped with cryogenic receivers. The space telescopes, observing together with ground-based radio telescopes, enable the highest angular resolution of 20 micro-arcsecond ($\mu$as) at 43 GHz. The SMVA is expected to conduct a broad range of high-resolution observational research, e.g. imaging the shadow (dark region) of the supermassive black hole in the heart of the galaxy M87 for the first time, studying the kinematics of water megamasers surrounding the SMBHs, and exploring the power source of active galactic nuclei. Pre-research funding has been granted by the CAS in October 2012, to support scientific and technical feasibility studies. These studies also include the manufacturing of a prototype of the deployable 10-m space-based telescope and a 22-GHz receiver. Here we report on the latest progress of the SMVA project.
\end{abstract}



\section{Introduction}
\label{intro}

The advent of the VLBI technique at the end of the 1960Õs has allowed astronomers on the Earth to observe the fine structure of compact celestial objects with angular resolution significantly surpassing that of optical telescopes. Exploiting baselines to a few thousand kilometers in length between the radio telescopes of a ground-based VLBI network, celestial objects and their environs can be imaged as if CCD sensors are used to photograph them through optical lenses. However, various parts and physical phenomena of a celestial object may be apparent at different wavebands, making it necessary to build diverse types of telescopes (radio, infrared, optical, X-ray, gamma-ray) to observe different astronomical features. 
The performance of a telescope is characterized by two major factors: resolution and sensitivity. For VLBI, these are equivalent to baseline length (resolution) and detection limit (sensitivity). Since the imaging resolution of VLBI is inversely proportional to the baseline length and is proportional to the observing wavelength, it is desirable to spread the VLBI telescopes over far-apart locations on the Earth, and to operate them at the highest possible frequency suitable for the intended applications. However, the baseline lengths of ground-based VLBI networks are constrained by the size of the Earth, while very-high-frequency radio radiation from the astronomical objects may be absorbed or blocked by the atmosphere of the Earth. For example, the Very Long Baseline Array  (VLBA) of the US \citep{Nap94} operates with baselines shorter than 8600 km and at frequencies at or below 86 GHz, resulting in the highest resolution of about 75 micro-arcsecond ($\mu$as). 

In order to break through the baseline length limit, astronomers must try to move the radio telescopes into space and allow them to fly in high-altitude orbits. This is known as Space VLBI. But clearly, such a space-VLBI approach is not a solution for every problem, as it also involves several serious technical difficulties for the payloads, such as building a large deployable space-based telescope with a diameter of 10-m or larger, the cooling system of the microwave equipment (to suppress the noise in the receiver amplifier), a highly stable time/frequency standard, high-rate data transfer to ground stations, high-accuracy pointing performance of the space antenna (a stringent requirement for the high efficiency of the telescope).

\subsection{History of Space VLBI}

The first ground-space VLBI experiment was successfully conducted by the TDRSS satellite in 1986 \citep{Lin90}. Since then, radio astronomers have been continuously attempting  to build and operate space VLBI telescopes with ground stations \citep{Gur13} Ð sometimes successfully, sometimes not. Interesting results have been obtained from some early efforts, and more are expected from a current mission named RadioAstron  led by Russia \citep{Kar13}. 
As the first successful dedicated space VLBI mission, the Japanese VSOP project launching the HALCA satellite is recognized as a milestone in the history of space VLBI and also in radio astronomy \citep{Hir98}. The HALCA satellite carried a 8-m space-borne radio telescope working jointly with radio telescopes on the Earth to form unprecedented baselines nearly 3 times as long as the EarthÕs diameter. It has made a lot of important accomplishments, such as the detection of nearly $10^{14}$K brightness temperature of AGN far exceeding the theoretical limit ($10^{12}$K) \citep{Frey00}, observation of hydroxyl (OH) masers at 1.6 GHz with extremely high brightness temperature \citep{Sly01}, detection of interference fringes and obtaining high-resolution VLBI images of quasars and radio galaxies etc \citep{Dod08}. Unfortunately HALCA could only be operated at 1.6 and 5 GHz (although it was also designed for 22 GHz), and the angular resolution was constrained by the its apogee altitude of 21400 km, which resulted in a limited angular resolution but usually a sparse (u,v) coverage between the Earth and space \citep{Hir00}. 
All things considered, VSOP is widely recognized as a rather successful mission despite some shortcomings mentioned above. Its greatest accomplishment is the real demonstration that high-resolution VLBI observations in a space-ground configuration with baselines longer than the diameter of the Earth are technically feasible and scientifically rewarding. The success of VSOP motivated scientists worldwide to explore the possibilities of launching more ambitious space VLBI missions in the future, which will benefit from possibly a range of new advanced electronic, mechanical and aerospace technologies. 
VSOP-2 is the follow-up project of VSOP, planned by the Japanese space agency JAXA as its second-generation space VLBI mission, featuring a maximum observing frequency of 43 GHz, a 10-m on-board antenna, and an apogee altitude of 26000 km. That would offer a greatly improved angular resolution of 38 $\mu$as \citep{VSOP2-jp,VSOP2-en,VSOP2-sci}. 
In the meantime, the US space agency NASA had also proposed a space VLBI mission called ARISE \citep{Ulv98} which, for budgetary reasons, was later replaced by its potential participation in VSOP-2. But NASA later promoted a dual-satellite mission known as iARISE, which featured an apogee altitude of 90 000 km corresponding to a further improved resolution of $\sim$7.5 $\mu$as, 15-m space antennas, and a sensitivity of $\sim$80 mJy at the highest observing frequency of 86 GHz \citep{Mur05,iARISE}. Unfortunately, both iARISE and VSOP-2 were postponed or cancelled. 
At the same time when the VSOP concept was first proposed, Russia also had a space VLBI project (named RadioAstron) that had been under active planning, design and construction for more than a decade. The mission finally succeeded in 2011. The satellite (Spektr-R) carrying a 10-m radio telescope was launched in July 2011 into a highly elliptical orbit with a maximal apogee altitude of $\sim$350000 km. Its highest observing frequency is 22 GHz, corresponding to an angular resolution of $\sim$7$\mu$as, and its baseline sensitivity is estimated to be $\sim$13 mJy \citep{RA}. Two years after the launch of RadioAstron, the radio astronomy community is eagerly looking forward to its important science findings. 
In the following section, we will introduce the active Chinese participation in the exciting field of space VLBI as a natural extension of its ground-based VLBI activities started in the late 1970s, and present the multi-stage roadmap of space VLBI envisioned for the next two decades. Much of the discussion will be focused on the design and scientific objectives of the first phase, {\it i.e.} the long-millimeter-wavelength space VLBI project.

\subsection{Roadmap of Chinese space VLBI}

China started elaborating the concept of its VLBI network already in the 1970Õs. Its first radio telescope was built in Shanghai with a diameter of 25 m in  1986, and then in Urumqi (also 25 m) in 1993. Later, in 2006, two new larger radio telescopes of 50-m and 40-m diameters were successively built in Beijing and Kunming. Starting in the 2010Õs, some very large radio telescopes began to emerge, including the 65-m radio telescope in Shanghai (commissioning in 2014) and the Five-hundred-meter Aperture Spherical Telescope (FAST) in Guizhou Province (to be completed in 2016). 
As radio astronomers in China were accumulating experience in building, operating and using ground-based radio telescopes for VLBI research over the past three decades, they were also gradually shifting their attention from the ground to space. A proposal for future "Space Millimeter-wavelength VLBI Array" (SMVA) was originally drawn in 2003, and was selected for pre-research in 2009 by the Space Science Project Committee of the Chinese Academy of Sciences (CAS). The pre-research project was then approved by the CAS in 2012 as a "Background Prototype Research," with the goal of completing the overall design of the first space VLBI array in the world within three years (2012-2015). It should be noted that this project, while technically challenging, must be science-driven with minimal risk. 
There will be a review meeting in 2015 to evaluate the progress made by that time, and to decide whether or not the space VLBI project should be selected as an engineering project. Below is a tentative roadmap for ChinaÕs space VLBI activities planned for the next two decades: 
\begin{itemize}
\item Stage 1 [2015-2020 planned]: Long mm-wavelength Space VLBI Array 
	 \begin{itemize}
		\item Two space telescopes (antenna diameter of 10 m) 
		\item Observing frequencies: 8, 22 and 43 GHz (7 mm)
		\item Highest resolution: 20 $\mu$as 
		\item Imaging capability better than ever
	\end{itemize}
\item Stage 2 [2021-2025]: mm-wavelength Space VLBI Array 
	\begin{itemize}
	\item Three space telescopes (12--15 m antennas) 
	\item Highest frequency: 86 GHz (3 mm)
	\end{itemize}
\item Stage 3 [after 2026]: Sub-mm-wavelength Space VLBI Array 
	\begin{itemize}
	\item Three to four space telescopes (12--15 m) 
	\item Sub-mm wavelength ($<$1mm)
	\end{itemize}
\end{itemize}

The following text will be focused on the scientific objectives, mission design and key techniques to be studied in Stage 1. 

\section{Scientific objectives}

The primary objective of the prospective Space VLBI Array is to substantially deepen and broaden our understanding of supermassive black holes (SMBHs) and the active galactic nuclei (AGN) in which they lie. Our Space VLBI satellite will operate in tandem with a number of ground radio telescopes in China and abroad to form baselines many times longer than the EarthÕs diameter, thus realizing an angular resolution as fine as 20 $\mu$as Ñ about 7.5 times of what can be achieved with ground-based interferometers (e.g. VLBA at 43 GHz) alone, or more than 20 times better than the decommissioned VSOP at 5 GHz. Observing at the highest frequency of 43 GHz, the proposed Space VLBI Array would be the only astronomical tool from 2020 onward which could directly image SMBHs and the hearts of AGN, measure their key physical properties, and track their evolution. 

\subsection{The nearest supermassive black hole -- M87}

The longest baseline ($>$60000 km) between the Space VLBI satellite and collaborating ground stations provides a high angular resolution of $\sim$20 $\mu$as (at 43 GHz). That would support: 
\begin{itemize}
\item Unveiling the emission structures surrounding SMBHs (mainly M87); 
\item Direct detection and imaging of the shadow (dark region) of M87.  
\end{itemize}

Direct imaging of a black hole is widely considered as a very rewarding endeavor \citep{BL09}, since the existence of BHs has been hitherto based on indirect evidence. But such a mission is also extremely challenging technically Ñ even if we are only zooming in on the more easily detectable SMBHs Ñ as the task requires state-of-the-art radio receivers, antennas, supporting electronics and cryogenic facilities to work cooperatively in a grand-scale interferometry network, offering sufficiently high angular resolution and sensitivity. Our prospective Space VLBI Array will be built to achieve this ambitious goal.

\begin{figure}
\includegraphics[]{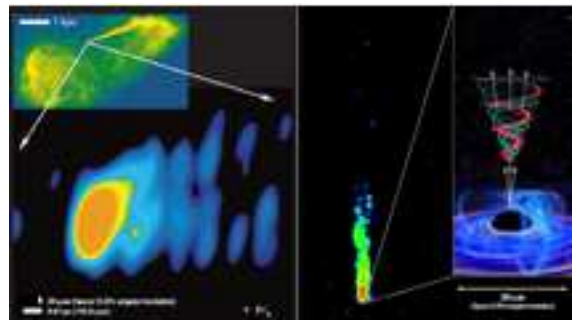}
\caption{Images of M87 obtained with ground VLBI stations (Left: \citep{Jun99,Hada11}) and a VSOP-ground configuration (Right), at very low resolutions in comparison with the 20-$\mu$as resolution of the proposed Space VLBI Array. (Image courtesy of NASA, NRAO).}
\end{figure}

The most appropriate candidate is M87, an elliptical galaxy harboring one of the nearest and most massive BHs, at a distance of 14.7 megaparsec (53.5 mega-lightyear). Hubble Space Telescope spectroscopy of its nucleus gave strong evidence for a rapidly rotating ionized gas disk at its center, from which the presence of a central SMBH with $\sim 3.2\times 10^9$ solar mass ($M_\odot$) was inferred. Its apparent shadow size was estimated to be $\sim26 \mu$as, which is comparable to the angular resolution of the Space VLBI Array, thus making it currently the best candidate for direct black hole imaging. Figure 1 shows the radio images of M87. The one in the bottom-left panel is obtained by several VLBI ground stations at 43 GHz ($\sim$150 $\mu$as resolution) \citep{Jun99}, and the other one in middle is obtained by VSOP in tandem with a ground station at 1.6 GHz ($\sim$1.0 mas resolution). The right panel displays an artistÕs impression of a BH. The comparison of the images on different scales clearly indicate that the expected higher resolution ($\sim$20 $\mu$as) of our prospective Space VLBI Array would be sufficient to resolve some interesting, yet-to-be-discovered details about the black hole and its environs. Examples for future study include the innermost portions of the accretion disk, the region in which material leaves the disk and spirals towards the SMBH event horizon, and the acceleration and collimation of ultra-hot plasma to form relativistic jet. 

\subsection{Supermassive black holes -- Megamasers}

\begin{figure}
\includegraphics[]{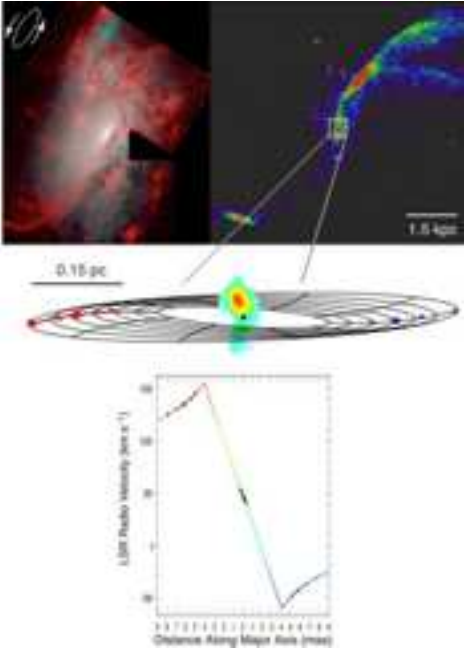}
\caption{Water megamasers in the accretion disk surrounding the SMBH in NGC4258 (\citep{Miy95,Her99}). Image courtesy of NRAO/AUI and L. Greenhill (Harvard-Smithsonian Center for Astrophysics).}
\end{figure}

Through imaging extragalactic water megamasers that reside in the accretion disk surrounding the central SMBH with a high $\sim$20 $\mu$as resolution (at 43 GHz), the Space VLBI Array would enable: 
\begin{itemize}
\item Direct mapping of the accretion disk structure, exploring the dynamics of accretion disk, and providing an accurate determination of SMBH mass. 
\item Determination of the Hubble constant H$_0$ and related extragalactic distances around the black holes by a proposed "MCP (Megamaser Cosmology Project) in Space" with sufficiently high resolution, accuracy, and sensitivity (e.g. $\sim$100 mJy in 100 kHz bandwidth at 43 GHz) required by the demanding task.
\end{itemize}

Besides directly imaging SMBHs (Section 2.1), another approach to gaining a deeper understanding of black holes is to explore extensively the accretion disks in AGN, as they are the key physical links between SMBHs and their host galaxies. Their proximity to the black holes makes them the best available probes of the deep gravitational wells and black hole environs. VLBI measurements of water maser positions, line-of-sight velocities and proper motions provide unique and direct, well-resolved maps of the underlying accretion disks (Figure 2). Ground-based VLBI networks with relatively poor angular resolutions could only study the nearest water megamasers, which are very limited in number. The Space VLBI Array has a definite advantage as it has a much higher angular resolution that could greatly extend the distance range of observations.

Observations will be carried out over a period of several years to possibly detect the proper motion and centripetal acceleration of mega-masers in several AGN. With these data one could detect rotation of the disks, which would allow for measuring the distances to galaxies geometrically, thus forming a robust basis for calibrating the extragalactic distance scale. 
The current MCP at NRAO \citep{MCP13} aims to determine the Hubble constant H$_0$ with high accuracy ($\sim$3\%) through measuring the angular-diameter distance to galaxies in the Hubble flow at distances of 50-200 Mpc. Better measurements of H0, which is related to the current expansion rate of the Universe, would provide critical independent constraints on dark energy, spatial curvature of the local Universe, neutrino physics, and the validity of general relativity. The MCP will also enable accurate determination of the central black hole mass in megamaser galaxies. At present, the MCP uses the GBT, VLBA, VLA, and Effelsberg radio telescopes to conduct the necessary observations. By comparison, the proposed Space VLBI Array is able to provide much longer baselines, hence higher angular resolutions and better accuracies. However, the practical challenge would be to overcome the sensitivity limit of the space-borne low-noise receiver and its cryogenic equipment. Currently the Space VLBI Array is expected to have a sensitivity of about 16.6 mJy (1$\sigma$, 512MHz bandwidth, 60s integration) when working in tandem with a 25-m VLBA telescope.

\begin{figure}
\includegraphics[]{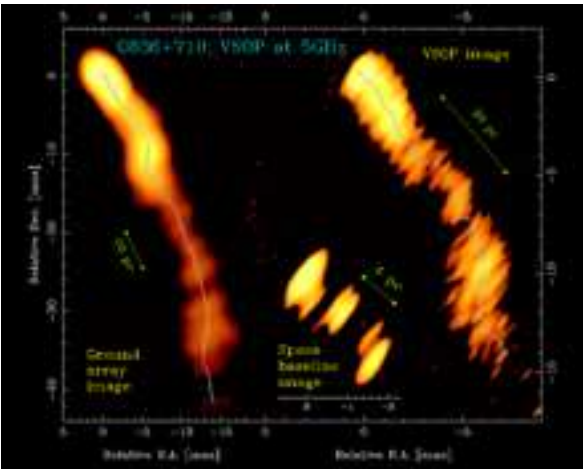}
\caption{Oscillation of the jet ridge line in 3C273 \citep{Lob01}. The right image is produced by the VSOP.}
\end{figure}

\subsection{Jets in active galactic nuclei}

To conduct a detailed study of the morphology, kinematics, and emission in extragalactic jets through probing a range of angular scales, these observations would help us to understand: 
\begin{itemize}
\item	Formation, acceleration, collimation of relativistic jets 
\item	Internal structure of jets 
\item	Polarization structure (B-field) 
\item	Origin of the high-energy X-ray and gamma-ray emissions 
\end{itemize}
Past VSOP observations revealed that AGN jets were extremely relativistic with the bulk Lorentz factors of $\Gamma \sim$ 30 \citep{Hir98}, but the formation, acceleration, and collimation mechanisms of the jets are still unclear.  The high-resolution VSOP image (Figure 3) reveals that the 3C273 jet shows a wiggling pattern of the ridge line, manifesting plasma instability in the jet flow \citep{Hada11}. The high-resolution polarized imaging capability of the proposed Space VLBI Array could enable a detailed study of these problems.

\subsection{Formation and evolution of massive stars}

Observing star-forming accretion disks and outflows traced by masers (H$_2$O at 22 GHz and SiO at 43 GHz) 
\begin{itemize}
\item	Again, sensitivity would be a big issue 
\item	Space VLBI would lead to deep exploration of maser physics in extreme situations 
\end{itemize}

With enhanced sensitivity resulting from better receivers, cryogenic equipment and high data rate, the ultralong space-ground baselines of the proposed Space VLBI Array will explore the star-forming regions on scales of 1-10 Astronomical Unit (1 AU = the mean distance between the Sun and the Earth). To date, dozens of 22-GHz H$_2$O masers have been found in locations near the sites where low-mass to high-mass stars were born. The gas kinematics associated with the birthplace of stars can be probed by H$_2$O masers using VLBI. On the other hand, SiO masers are formed in the outflows of late-type Asymptotic Giant Branch stars. SiO maser kinematics allow us to trace this important late stage of stellar evolution. Several SiO maser transitions fall in the 43 GHz observing band supported by the Space VLBI Array, and they are known to be bright and extremely compact. Imaging these maser sources helps to address a number of key scientific questions related to the late stellar evolution.

\section{Orbit design and (u,v) coverage}

Our prospective space VLBI array project underwent intensive in-house studies and extensive peer reviews in the past few years. So far we have identified some key mission drivers that could result in an overall satisfactory project with the "best" returns possible, after making a few necessary compromises to reach some optimal performance trade-offs. 

\begin{figure}
\includegraphics[scale=0.3]{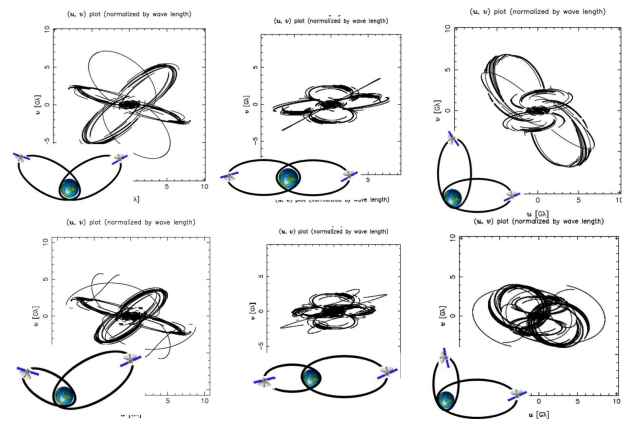}
\caption{Orbital design and resulting (u,v) coverage \citep{An13}. The orbit parameters are given in Table 1.}
\end{figure}

\begin{figure}
\includegraphics[scale=0.25]{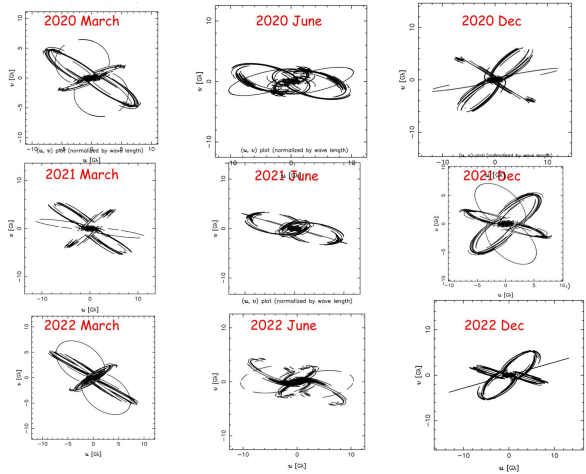}
\caption{Time evolution of the (u,v) coverage due to orbit precession using Orbit Configuration 1 as an example \citep{Waj13,An13}.}
\end{figure}

A key element of the mission design is the complementary dual-spacecraft orbit configuration that would give good (u,v) coverage not only for all sky, but also for specific relevant targets such as M87. Some possible configurations are presented in Table 1. Some simulation results of the (u,v) coverage for M87 are displayed in Figure 4 \citep{Waj13,An13}, showing good (u, v) coverage when the two satellites are working with ground stations worldwide. The co-planar orbits of two satellites are less attractive because the (u,v) points are concentrated in one direction. Orbits with unequal apogee heights (e.g., 60000 km and 40000 km) could be a good compromise between the longest and intermediate (u,v) coverage. Due to the orbit precession, the (u,v) coverage may change with time for a certain celestial source. Figure 5 shows an example of the time evolution of (u,v) coverage of Orbit Configuration 1.

The data observed at different dates can be accumulated to improve (u,v) coverage as long as the  timescale of intrinsic structural variation is less than the time span of the data sets. The orbit design could be further refined in the future based on comprehensive study of the brightness distribution resulted from the simulated visibilities. More practical evaluation should also include thermal noise and instrumental noise. 

\begin{table}
\small
\caption{Orbital parameters for six configurations shown in Figure 2.}
\begin{tabular}{ccccccc}
\hline
Model &  & Apogee & Perigee & $\theta$($^\circ$) & $\Omega$ ($^\circ$) & $\omega$ ($^\circ$) \\ \hline
1 & SC-1 & 60000 & 1200 & 28.5 & 15  & 220 \\
  & SC-2 & 60000 & 1200 & 28.5 & 180 & 150 \\ \hline
2 & SC-1 & 60000 & 1200	& 0	   & 15  & 220 \\ 
  & SC-2 & 60000 & 1200	& 180  & 180 & 150 \\ \hline
3 & SC-1 & 60000 & 1200	& 0    & 15	 & 220 \\
  & SC-2 & 60000 & 1200	& 90   & 180 & 150 \\ \hline
4 & SC-1 & 60000 & 1200	& 28.5 & 15  & 220 \\
  & SC-2 & 40000 & 1200	& 28.5 & 180 & 150 \\ \hline
5 & SC-1 & 60000 & 1200 & 0    & 15  & 220 \\
  & SC-2 & 40000 & 1200 & 180  & 180 & 150 \\ \hline
6 & SC-1 & 60000 & 1200 & 0    & 15  & 220 \\
  & SC-2 & 40000 & 1200 & 90   & 180 & 150 \\ \hline
\end{tabular} \\
Note--- Column 3-4: apogee and perigee height; column 5: inclination angle of the satellite orbital plane; column 6: Right Ascension of ascending node means RA direction of orbital plane; column 7: argument of perigee means in which direction the orbit is elongated.
\end{table}

\section{Key technologies}

The ultimate success of our prospective Space VLBI Array depends on a number of key technologies that would enable all the components/subsystems on the satellite and in the ground stations to operate properly together as a system, performing the missionÕs required functions according to strict specifications. At present, after several rounds of initial designs, experimentation, reviews and revisions, we have identified four key technologies (described below) for more focused development. However, we must emphasize that this shortlist may likely be refined or substantially changed as we are making more progress in the next few years. 

\begin{description}
\item[Space Antenna:] The satellite will have a 10-m Casse-grain-type deployable antenna with a total mass $\leq$400 kg, whose reflector will have a surface error $\leq$0.4 mm (RMS) as required by the 43-GHz observation. When folded, the antenna should fit into a cylinder with a 3.1-m diameter and 5.1-m height. To meet the stringent surface error requirement, special alloys will be used to reduce possible deformation of the mesh structure in the reflector due to significant day/night temperature changes when operating in space, and probably because of the dew caused by air humidity during launch as well. 
\item[Data Receiving and Acquisition System:] The space antenna will operate in three bands, namely, 6-9 GHz (X) with efficiency $\geq 60\%$; 20-24 GHz (K) with efficiency $\geq 50\%$; and 40-46 GHz (Q) with efficiency $\geq 40\%$. The receivers will support LHCP/RHCP dual polarizations, and the low-noise amplifiers will be cooled down to about 30 K with cryogenic facilities to reduce its instrumental noise when operating at 22/43 GHz. 
\item[Antenna Pointing and Satellite Attitude Control: ] The space antenna will have a pointing error $\mu$15 $\mu$as or 0.0042$^\circ$ (1$\sigma$). Since the antenna is physically tied to the satellite platform, the antenna pointing is controlled by changing the satelliteÕs attitude. Hence the satellite attitude control should also have a pointing error $\leq15 \mu$as. (Note: Since phase-referencing technique will not be used to increase the integration times of the detected signals with techniques like quick "nodding", the rate of changing the antennaÕs pointing angle is not a serious issue to consider here.) 
\item[Time and Frequency Standard:] To support observing at 43 GHz, the time/frequency standard on board the satellite should have relatively good short- and long-term stabilities, comparable to those of H-masers with the following tentative specifications: $3 \times 10^{-12}$ (1 sec) and $1 \times 10^{-14}$ (daily). To meet these requirements, a space-qualified H-maser is considered to be installed on the satellite. On the other hand, a mechanism that will effectively remove frequency shifts due to the satelliteÕs Doppler (with respect to the ground stations) will also be provided to "transfer" the highly stable phase and frequency of an atomic clock on the ground to the satellite, so that an on-board USO (ultra-stable-oscillator) can be "disciplined" to accurately follow the phase and frequency of the ground-based atomic clock. This two-pronged approach may help to reduce the risk of on-board H-maser failures, and also to solve Ñ at least partially Ñ the problem of having only a limited number of collaborating up-link tracking stations worldwide. 
\end{description}

Other less crucial, but also important, technologies to develop include: 
\begin{itemize}
\item	High-rate data communication from the spacecraft to the ground receiving stations at 1/2 Gbps. For dual-polarization observation, 1 Gbps will only allow a bandwidth of 256 MHz, while 2 Gbps will accommodate a preferred 512 MHz bandwidth. Our objective is to provide the highest possible data rates while maintaining acceptable bit-error-rates for given S/N ratios in the data link (under all possible weather conditions) that may be as long as 60000 km. 
\item	A laser reflector on the satellite may be installed to enable accurate Doppler measurement and orbit determination by ground tracking stations. 
\item	A high-volume mass storage on the satellite may be installed to buffer the captured astronomical data for some periods of time (probably a few hours) when no ground tracking stations are visible.
\end{itemize}

\section{Conclusions}
The Space Millimeter VLBI Array presented here is the most important first mission of ChinaÕs current multi-stage roadmap that would ultimately lead to a sub-mm-wavelength space VLBI network enabling high-resolution ($\mu$as or better) observation. This mission would also complement other possible space telescopes observing in the X-rays or ground-based submm-wavelength VLBI within the next 10-20 years. 
Although challenging in technique, the risks of the proposed science-driven mission could be reduced to a minimum by using advanced but mature technologies that might be further improved in the next few years. If the two satellites of this project could be launched and put into operation around 2020 as planned, it would form the worldÕs first space VLBI network working at 43 GHz. The ultrahigh (20 $\mu$as) resolution and the outstanding imaging capacity enable direct imaging of supermassive black holes, the central engines of AGN and other compact radio sources. Hence the returns of this mission would be very high as such unique capabilities in astronomical observation could substantially broaden our knowledge and deepen our understanding of the Universe.

\section*{Acknowledgements}
This project is supported by the Strategic Priority Research Program on Space Science of the Chinese Academy of Sciences (grant No. XDA04060700). The authors thank Leonid Gurvits, Hisashi Hirabayashi, Ken Kellermann, Yoshiharu Asaki, Willem Baan and Sandor Frey for helpful discussion.

\label{lastpage}

\end{document}